\documentclass[journal]{IEEEtran}
\usepackage{amsmath,epsfig,epsf,times,amssymb,booktabs,subfigure}
\usepackage{setspace}
\usepackage{array}
\usepackage{url}
\usepackage{float}
\usepackage{color}
\usepackage{graphicx}
\usepackage{amsthm,cite}
\usepackage{enumerate}
\usepackage{cite}
\usepackage{multirow}
\usepackage{balance}
\theoremstyle{remark}

\begin{document}

\title{Physical Layer Security for Ultra-Reliable and Low-Latency Communications}
\author{Riqing Chen, Chunhui Li, Shihao Yan, Robert Malaney, and Jinhong Yuan
\thanks{R. Chen is with Fujian Agriculture and Forestry University; C. Li is with the Australian National University; S. Yan is with Macquarie University; R. Malaney and J. Yuan are with the University of New South Wales. The corresponding author of this article is Shihao Yan.}
\thanks{This work was supported in part by the National Natural Science Foundation of China (Grant No. 61501120, No. 61603095, No. 71601049), the Digital Fujian Institute of Big Data for Agriculture and Forestry (Grant No. KJG18019A), and the Australian Research Council under Discovery Project Grant (DP180104062).}}

\maketitle

\vspace{-1cm}

\begin{abstract}
Ultra-reliable and low-latency communication (URLLC) is one category of service to be provided by next-generation wireless networks. Motivated by increasing security concerns in such networks, this article focuses on physical layer security (PLS) in the context of URLLC. The PLS technique mainly uses transmission designs based on the intrinsic randomness of the wireless medium to achieve secrecy. As such, PLS is of lower complexity and incurs less latency than traditional cryptography. In this article, we first introduce appropriate performance metrics for evaluating PLS in URLLC, illustrating the tradeoff between latency, reliability, and security. We  then identify the key challenging problems for achieving PLS for URLLC, and discuss the role that channel state information can have in providing potential solutions to these problems. Finally, we present our recommendations on future research directions in this emerging area.
\end{abstract}


\section{Introduction}

\par Next-generation wireless networks (fifth generation and beyond) will be fundamentally different from previous generations in many regards. Perhaps most importantly (and contrary to earlier  generations of wireless networks which focused on data rate as the single most important requirement), next-generation wireless networks will have ultra-reliable and low-latency communication (URLLC) as a  requirement.   URLLC is envisioned to enable the wireless exchange of data packets  with ultra-high reliability (error probability on the order of $10^{-7}$) and ultra-low latency (end-to-end delay on the order of 1 ms) \cite{LiPIEEE2018}. Such communication embodies a new wireless paradigm in which ``click-and-wait'' communications are replaced by dependable real-time interactive communications. In this new paradigm, the assurance  of ultra-high reliability  creates confidence that wireless communications can be used even in life-threatening circumstances, while ultra-low latency ensures real-time functionality in time-critical interactive communications.
The emergence of URLLC will enable many new time-critical applications such as autonomous networked vehicles, next-generation factory automation, tele-surgery and the Tactile internet \cite{ji2018ultra}, thereby  opening up  lucrative new business opportunities for many industrial sectors. As such, a large amount of research effort has been dedicated to URLLC in recent years. Such research efforts continue to grow, with an increasing resolve to find technical solutions to the somewhat contradictory requirements of ultra-high reliability and ultra-low latency.

\begin{figure}[t]
\centering
\includegraphics[width=0.48\textwidth]{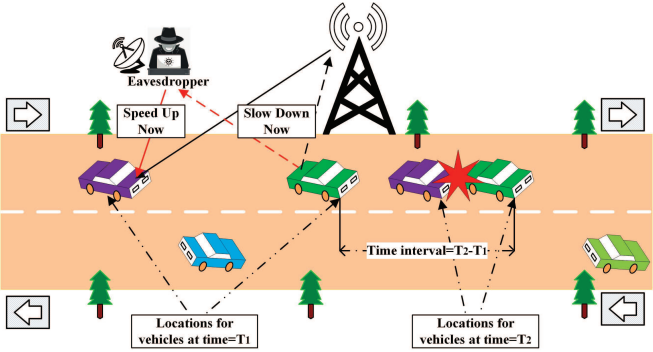}
\caption{An attack scenario in vehicular ad-hoc networks showing the criticalness of URLLC security.}
\label{fig:fig1}
\end{figure}

Although the key requirements of URLLC are mainly related to reliability and latency, security issues are also  critical in most application scenarios of URLLC. For example, in addition to ultra-low latency and ultra-high reliability, other key requirements arising from the nature of the Tactile internet are security and privacy \cite{LiPIEEE2018}. In addition, the leakage of critical and confidential information in some applications of URLLC may lead to attacks that are difficult to defend against.
{We present an example in Fig.~\ref{fig:fig1} to further demonstrate this point, where $T_1$ is the time when we have the scenario on the left-half figure and $T_2$ is the time when the scenario shown on the right-half figure. In this figure, we show that if an eavesdropper obtains a legitimate vehicle's message together with its random sequence (used for authentication) through eavesdropping, the eavesdropper can successfully replace the legitimate message (i.e., ``slow down'' sent by the green car in Fig.~\ref{fig:fig1}) with some misleading information (i.e., ``speed up'' sent by the eavesdropper in Fig.~\ref{fig:fig1}) within vehicular networks. Such misleading information can potentially result in fatal accidents~\cite{EltayebICC2018}.}
{We note that this example demonstrates the importance of both the information confidentiality and message integrity, since the leaked random sequence leads to the fake misleading information being accepted (message integrity is not guaranteed). A brief discussion on the importance of message integrity in the context of URLLC can be found in~\cite{chen2018ultra} and in this work we focus on the information confidentiality. We also note that information confidentiality is widely addressed by cryptography algorithms in traditional communications.}
{However, in the context of URLLC, cryptography algorithms may violate ultra-low latency requirements due to the high-complexity signal processing required by encryption and decryption~\cite{LiPIEEE2018}. In addition, the key distribution required by cryptograph solutions may cause extra delay in some application scenarios.}

Different from  traditional cryptography algorithms, physical layer security   mainly utilizes transmission techniques, and the inherent properties of the wireless medium, to achieve secrecy \cite{yang2015safe}. In the wiretap channel model for physical layer security, a transmitter, Alice, sends confidential information to a legitimate receiver, Bob, while an eavesdropper, Eve, attempts to interpret this confidential information by eavesdropping on Alice's transmission. One advantage of physical layer security techniques is that they do not require key distribution or encryption/decryption.
%which can significantly reduce the signal processing complexity and thus the communication latency.
Thus, physical layer security can potentially defend against eavesdropping attacks without violating the ultra-low latency requirement of URLLC. Going forward, therefore, physical layer security may well be the main security technology invoked to  protect URLLC in next-generation wireless  networks.

Physical layer security has previously been widely studied in wireless communications, but largely without consideration of an ultra-low latency or ultra-high reliability constraint.
As such, many fundamental questions about physical layer security in the context of URLLC remain unclear. For example, \textit{``What are the proper secrecy metrics used to evaluate physical layer security in URLLC scenarios?''} is just one of these questions. Due to the finite (and small) blocklengths considered in URLLC, the decoding error probabilities at both Bob and Eve are not negligible in wiretap channels. Another consequence of small blocklengths is that many widely-used secrecy metrics cannot be used to evaluate the performance of physical layer security in URLLC scenarios. For example, the well-known secrecy capacity and secrecy outage probability both require an infinite blocklength $n$.
In addition, the information-theoretic strong secrecy and weak secrecy in the context of physical layer security are both defined in the limit of $n \rightarrow \infty$, and therefore they also are not applicable to URLLC.

Channel state information (CSI) is required in many of the techniques  used to enhance the reliability and security of wireless communications. However, considering the short blocklengths required by the ultra-low latency in URLLC, there may not be the required amount of channel uses that would be needed for accurate channel estimation.
Against this background, many problems related to CSI in physical layer security for URLLC are challenging and should be revisited. The initial challenge is to determine whether (and when) CSI is required for achieving physical layer security in URLLC scenarios - a question that is hard to directly clarify  given the limited channel uses. We do note, however, that communication without the use of CSI may bring one benefit for physical layer security in URLLC in that it potentially  limits Eve's channel estimation  ability. This could force Eve to use non-optimal techniques for eavesdropping on Alice's confidential information.

In this article we  discuss all of the above issues in some detail - organized as follows. In Section~\ref{sec:metrics}, we review the used secrecy performance metrics in the literature and clarify whether they are applicable to physical layer security in URLLC scenarios. Challenging problems related to CSI and potential solutions for achieving and enhancing physical layer security for URLLC are presented in Section~\ref{sec:CSI}. We conclude this article in Section~\ref{sec:conclusion}.

\section{Performance Metrics for Physical Layer Security in URLLC}\label{sec:metrics}

In this section, we review the widely used secrecy performance metrics for physical layer security and clarify whether they can be adopted to quantify the achievable physical layer security in URLLC scenarios.

\subsection{Information-Theoretic Secrecy Performance Metrics}\label{Sec:internal-weak}

Perfect secrecy was initially used as the secrecy coding metric by Shannon, which requires zero mutual information between $M$ and $X^n$, where $M$ is the transmitted message and $X^n$ is a length-$n$ codeword used to transmit $M$ \cite{HarrisonSPM2013}. We note that  `zero' mutual information means that the coded message $X^n$ does not provide any information on the message $M$. Perfect secrecy can also be taken to mean the mutual information between $M$ and $Z^n$ is zero, where $Z^n$ is the received symbol at Eve. This means that Eve cannot obtain any information on the transmitted message $M$ from her received symbol $Z^n$. The definition of  perfect secrecy is valid for arbitrary values of the codeword length $n$. As such, it is suitable for defining secrecy in physical layer security for URLLC. However, as Shannon proved,  perfect secrecy can only be achieved when the entropy of the secret key is  at least that of the entropy of the transmitted message itself - leading to the fact that the notion of perfect secrecy is impractical. Therefore, we can conclude that the perfect secrecy is not a practical metric for evaluating physical layer security in URLLC.

Considering the impractical issues of perfect secrecy, weak secrecy and strong secrecy were proposed as alternative secrecy metrics for  physical layer security. Weak secrecy is achieved if the per-channel use mutual information between $M$ and $Z^n$ approaches zero as $n$ approaches infinity, while strong secrecy is achieved if the total mutual information for all channel uses approaches zero as $n$ becomes infinite~\cite{HarrisonJWCN2018}.
Although the weak and strong secrecies are achievable by practical coding strategies \cite{HarrisonSPM2013}, as per their definitions, it is hard to evaluate their uses in URLLC in the context of physical layer security. This is again due to the fact that their definitions require $n$ being infinite, while in URLLC $n$ must be finite and small due to the required ultra-low latency.

In  physical layer security, secrecy capacity is another widely used secrecy metric. In most existing works with secrecy capacity as the performance metric, it has not been clarified whether its definition is based on  weak secrecy or strong secrecy. We would like to clarify that  secrecy capacity is not a proper performance metric for physical layer security in URLLC, no matter whether it is defined based on weak secrecy or strong secrecy \cite{HarrisonJWCN2018}.
{For example, the widely used secrecy capacity is defined as the supremum of the code rate (i.e., the information rate) that can} achieve weak secrecy (against a passive Eve) as a function of wiretap channel parameters, while guaranteeing an arbitrarily low error probability at Bob \cite{HarrisonJWCN2018}.
As such, this secrecy capacity cannot be used as a metric for a finite small $n$ in URLLC, due to the fact that weak secrecy requires $n \rightarrow \infty$.
Intuitively, this can be explained by the fact that the main channel capacity $C_b$ or the eavesdropper's channel capacity $C_e$ cannot be achieved with arbitrarily low error probabilities.

\begin{figure}[t]
\centering
\includegraphics[width=0.48\textwidth]{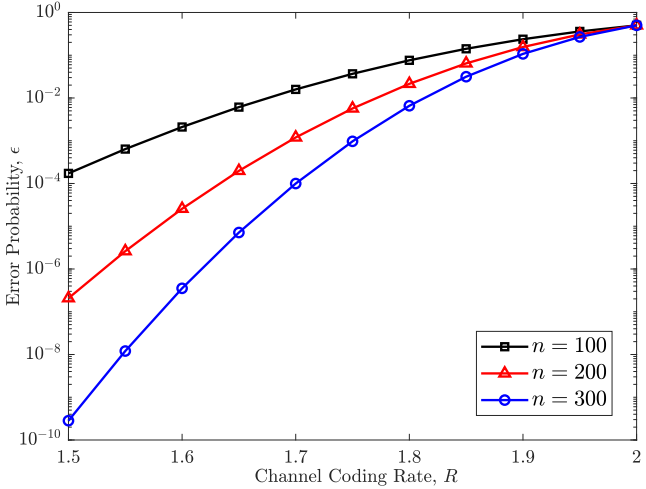}
\caption{Error probability $\epsilon$ versus channel coding rate $R$ for different values of the finite blocklength $n$.}
\label{fig:fig2}
\end{figure}

Due to the same reasoning as outlined above, another widely used secrecy metric, i.e., secrecy outage probability, also cannot be used to evaluate physical layer security in URLLC scenarios. For a finite blocklength $n$, there exists a tradeoff between the channel coding rate $R$ and the corresponding error probability $\epsilon$ \cite{poly2010channel}. As $n \rightarrow \infty$, this channel coding rate approaches the channel capacity while $\epsilon \rightarrow 0$. To further illustrate this point, in Fig.~\ref{fig:fig2} we plot the error probability $\epsilon$ (its expression is taken from \cite{poly2010channel}) versus the channel coding rate $R$ for different values of the blocklength $n$.
In addition to the tradeoff between $R$ and $\epsilon$, in this figure we observe that for a fixed $R$, $\epsilon$ increases significantly as $n$ decreases in the low regime of $n$. This demonstrates that the error probability $\epsilon$ is not negligible in URLLC. For physical layer security in URLLC, this non-negligible error probability exists at both Bob and Eve. As such, a fair secrecy performance metric should consider the impact of $R$, $\epsilon$, and $n$ in URLLC scenarios.

\subsection{Non-Information-Theoretic Secrecy Performance Metrics}\label{Sec:internal-strong}

Considering specific coding schemes (e.g., low-density parity-check codes), the authors of \cite{KlincTIFS2011} proposed to use the so-called security gap as a measure of secrecy - a metric based on the average bit-error rate (BER). The security gap is defined as $\text{SNR}_{B, min}/\text{SNR}_{E, max}$, where $\text{SNR}_{B, min}$ denotes the reliability threshold and $\text{SNR}_{E, max}$ denotes the security threshold. The reliability threshold is defined as the lowest signal-to-noise ratio (SNR) of the main channel that ensures the reliability requirement that $P_{\text{BER}}^B$ is not higher than $P_{\text{BER}, max}^B$. Here, $P_{\text{BER}}^B$ is the average BER at Bob, and $P_{\text{BER}, max}^B (\approx0)$ denotes the maximum average BER required to guarantee the reliability of the communication from Alice to Bob.
The security threshold is defined as the highest SNR of the eavesdropper's channel that ensures the security requirement that $P_{\text{BER}}^E$ is higher than $P_{\text{BER}, min}^E$, where $P_{\text{BER}}^E$ is the average BER at Eve and $P_{\text{BER}, min}^E (\approx0.5)$ denotes the minimum average BER that guarantees a certain level of secrecy (e.g. Eve cannot extract much information on Alice's transmitted message from her received signals). Considering that the average BER can be evaluated for arbitrary values of the blocklength $n$, the proposed security gap can be used to evaluate physical layer security in URLLC.

Also based on the average BER, the rate interval was proposed as another secrecy metric to evaluate physical layer security with a finite blocklength $n$~\cite{KimTCOM2016}. Mathematically, the rate interval is given by $\Delta R = R_{\text{sup}} - R_{\text{inf}}$, where $R_{\text{sup}}$ is the highest allowable transmission rate to satisfy $P_{\text{BER}}^B$ being less than or equal to  $P_{\text{BER}, max}^B$ and $R_{\text{inf}}$ is the lowest allowable transmission rate to satisfy $P_{\text{BER}}^E$ being larger than or equal to $P_{\text{BER}, min}^E$. As clarified in \cite{KimTCOM2016}, the rate interval is not always positive. When $\Delta R$ is positive, Alice is able to transmit information reliably and securely to Bob.
When $\Delta R$ is negative, it is not possible for Alice to set a transmission rate such that the above two constraints are satisfied simultaneously and thus Alice's reliable and secure information transmission should be suspended. We note that this rate interval converges to the secrecy capacity as $n$ approaches infinity. This is due to the fact that, in the limit of $n \rightarrow \infty$, the main channel capacity $C_b$ is achievable with an arbitrarily low average BER and Eve cannot obtain any information on Alice's transmission as long as the transmission rate is greater than $C_e$. For a finite $n$, as clarified in \cite{KimTCOM2016}, $R_{\text{sup}}$ is visibly lower than $C_b$, but $R_{\text{inf}}$ cannot be significantly lower than $C_e$. As such, we may still have the rate interval being negative even when the secrecy capacity is positive. This demonstrates the tradeoff among communication latency, reliability, and security in the context of URLLC, and indicates that the rate interval  can be used as a performance metric for physical layer security in URLLC.

Determining the expression of the average BER requires a specified  coding scheme. However, general expressions for the average BER are hard to obtain for coded systems. This is one issue that works against the aforementioned security gap and rate interval metrics in the context of URLLC. In order to overcome this issue, the bit-error cumulative distribution function (BE-CDF) and bit-error rate cumulative distribution function (BER-CDF) were proposed as alternative metrics to replace the average BER for defining the  security gap and rate interval~\cite{HarrisonJWCN2018}. The calculations of the BE-CDF and BER-CDF only require the error correction capability of a coding scheme (which is relatively easier than the calculation of the average BER), and thus lead to a tractable performance analysis on physical layer security with a finite $n$. In addition, as clarified in \cite{HarrisonJWCN2018}, the BE-CDF and BER-CDF provide more information on the BER performance of a system and thus they can offer a stronger secrecy guarantee for URLLC. 

%Following \cite{HarrisonJWCN2018}, we present Table~I to summarize our main points of the discussed various secrecy performance metrics.

In  physical layer security, another secrecy requirement given a finite blocklength $n$ is that based on the average probability of decoding error \cite{RajeshICCS2012}. In general, this secrecy requirement can be represented by the reliability constraint that $P_e^n(B)$ is not higher than $\beta_1$, and the security constraint that $P_e^n(E)$ is not lower than $\beta_2$, where $P_e^n(B)$ and $P_e^n(E)$ denote the average probabilities of decoding error at Bob and Eve, respectively. Here, $\beta_1$ is the maximum allowable average probability of decoding error at Bob, and $\beta_2$ is  a lower bound  on the average probability of decoding error at Eve.
As mentioned in \cite{RajeshICCS2012}, the blocklength $n$ should be sufficiently large in order to simultaneously guarantee the reliability constraint and the security constraint. As such, these two constraints may not be simultaneously satisfied for a finite small $n$ (this is similar to the fact that the rate interval based BER metrics may not be always positive). This again explicitly shows the tradeoff among latency, reliability, and security in wireless communications~\cite{YangISIT2017}.

\begin{figure}[t]
\centering
\includegraphics[width=0.48\textwidth]{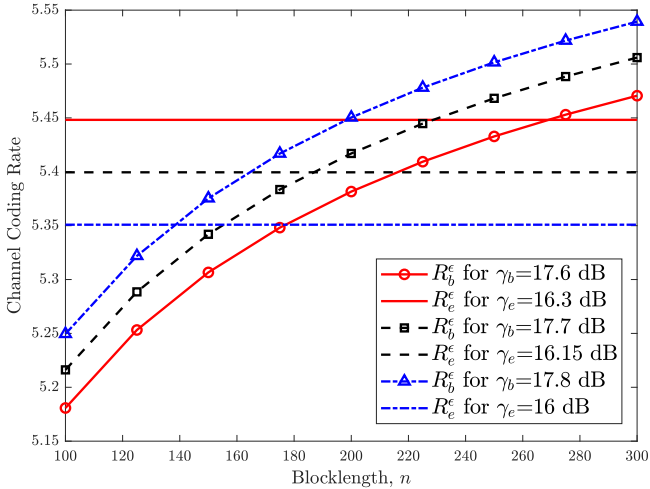}
\caption{Maximum main channel coding rate $R_b^{\epsilon}$ and minimum eavesdropper's channel coding rate $R_e^{\epsilon}$ versus the blocklength $n$ for different values of Bob's SNR $\gamma_b$ and Eve's SNR $\gamma_e$, where $\beta_b = 10^{-6}$ and $\beta_e = 0.5$.}
\label{fig:fig3}
\end{figure}

In order to further demonstrate the tradeoff among latency, reliability, and security in URLLC scenarios, we plot Fig.~\ref{fig:fig3} based on the research work of \cite{poly2010channel}. To this end, we use the error probability given in \cite{poly2010channel} to replace the previously discussed average BER, BE-CDF, BER-CDF, and the average probability of decoding error. The calculation of the error probability in \cite{poly2010channel} does not require a particular coding scheme, the error correction capability of a coding scheme, or the modulation method, since this specific error probability approximates the error probability that can be achieved by any scheme. Accordingly, the secrecy requirement can be written as the reliability constraint ($\epsilon_B$ being not higher than $\beta_b$) and the security constraint ($\epsilon_E$ being not lower than $\beta_e$), where $\epsilon_B$ and $\epsilon_E$ are the error probabilities at Bob and Eve, respectively. The error probability $\epsilon_B$ monotonically increases with the main channel coding rate $R_b$ for fixed $n$, and with the SNR at Bob $\gamma_b$ \cite{sun2018noma}. This leads to the reliability constraint determining an upper bound $R_b^{\epsilon}$ on $R_b$, for fixed $n$ and $\gamma_b$. Likewise, the security constraint determines a lower bound $R_e^{\epsilon}$ on the eavesdropper's channel coding rate $R_e$ (for fixed $n$) and the SNR at Eve $\gamma_e$. We note that $R_b^{\epsilon}$ and $R_e^{\epsilon}$ correspond to $R_{\text{sup}}$ and $R_{\text{inf}}$ in the rate interval metric that is defined using BER metrics.
In Fig.~\ref{fig:fig3}, we plot $R_b^{\epsilon}$ and $R_e^{\epsilon}$ versus the blocklength $n$ for different values of $\gamma_b$ and $\gamma_e$. We note that the reliability and security constraints can only be simultaneously satisfied when $R_b^{\epsilon}$ is not lower than $R_e^{\epsilon}$.
{We also note that the curves for $R_e^{\epsilon}$ are horizontal in Fig.~\ref{fig:fig3}. This is due to $\beta_e = 0.5$, for which $R_e^{\epsilon}$ is the same as the corresponding channel capacity, which is not a function of the blocklength $n$.}
In this figure, we also observe that $R_b^{\epsilon}$, being not lower than  $R_e^{\epsilon}$, is more likely to be satisfied for a larger $n$ in terms of requiring a smaller value of $\gamma_b/ \gamma_e$ (which is one type of the security gap). This once again explicitly demonstrates the tradeoff among latency, reliability, and security, and indicates that the security gap and rate interval based on error probability are valid performance metrics for physical layer security in URLLC scenarios. We note that the secrecy metrics discussed in this subsection are non-information-theoretic, which leads to the fact that the security measured by these metrics are not information-theoretically guaranteed. However, such security is acceptable in URLCC scenarios, since the ultra-low latency forces that the outdated confidential information is of no value~\cite{KlincTIFS2011}.

\begin{figure}[t]
\centering
\includegraphics[width=0.48\textwidth]{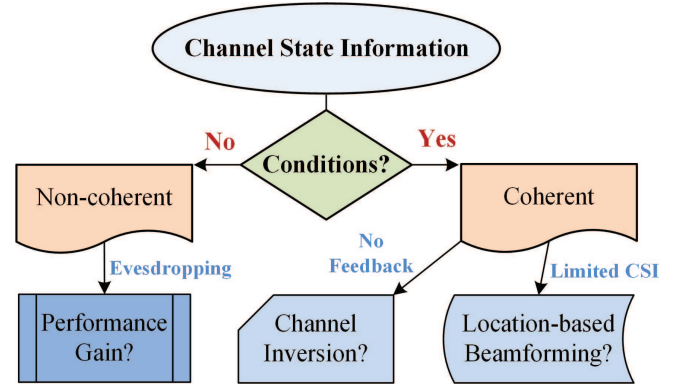}
\caption{Challenging problems and potential solutions on CSI issues for physical layer security in URLLC. Here 'Limited CSI' refers to imperfect CSI.}
\label{fig:fig4}
\end{figure}

\section{CSI Issues on Achieving Physical Layer Security in URLLC}\label{sec:CSI}

{Current research on physical layer security for URLLC is in its preliminary stage and further exploration is required. As such, many challenging issues remain in the implementation of physical layer security in practical URLLC scenarios, including, but not limited to,  (i) wiretap coding to simultaneously meet the latency, reliability, and security requirements, (ii) designing efficient feedback strategies with optimal tradeoffs between quantization accuracy and overhead cost, and (iii) achieving accurate CSI under the constraints of ultra-low latency, ultra-high reliability, and moderate levels of security. As we clarified previously, CSI plays a critical role in achieving physical layer security in URLLC, since CSI is required in many techniques used to enhance wireless communication reliability and security (e.g., beamforming and artificial-noise-aided secure transmission schemes). Therefore, in this section, we first clarify some design challenges related to CSI in achieving physical layer security in URLLC, and then  point out some potential techniques to overcome these challenges. The challenges we raise  are schematically  shown in Fig.~\ref{fig:fig4}.}

\subsection{Coherent Communications versus Non-Coherent Communications}

As mentioned in the Introduction, due to the required ultra-low latency of URLLC, there may not be enough channel uses to perform channel estimation and feedback during one communication block. This leads to the necessity of conducting  communications in URLLC where no CSI is estimated or fed back to a receiver before information transmission. We refer here to such communications as non-coherent communications. Actually, non-coherent communications bring one benefit for physical layer security, i.e., non-coherent communications do not require any estimation or feedback of the CSI of the main channel, thus limiting Eve's opportunity to  obtain the CSI of the eavesdropper's channel. This in turn reduces the eavesdropping capability of Eve by enforcing her to use non-coherent communication techniques.
{Meanwhile, we note that the unknown CSI also decreases the performance of the communication from Alice to Bob, i.e., non-coherence communication leads to a low reliability of the transmission from Alice to Bob. As such, how to enhance the reliability of non-coherent communications should be tackled before applying this technique into URLLC scenarios that attempt to achieve physical layer security. Therefore, non-coherent communications have a double-sided impact (reducing latency but also decreasing reliability) on physical layer security in URLLC relative to coherent (full CSI) communications. Accordingly, a significant future research direction for URLLC is to clarify the conditions under which non-coherent communications outperform coherent communications, and how much performance gain can be achieved by each in the context of physical layer security for URLLC.}

\subsection{Channel Inversion Power Control based on Channel Reciprocity}

In some applications (e.g., emergency alert systems), only uplink or downlink communications require URLLC. In such applications, channel inversion power control (CIPC) based on channel reciprocity \cite{elsawy2014on} can be used to overcome the CSI issues. We use the example where only the uplink from a user to a base station (BS) requires URLLC and security to demonstrate this point. Since the downlink communication does not require low latency, pilots can be periodically transmitted by the BS such that the user can estimate the downlink channel $h_d$. When the uplink communication with URLLC requirement is on demand, the user can use CIPC based on channel reciprocity (i.e., the uplink channel $h_u$ is the same as $h_d$) to guarantee that the power of the received signals is a constant $Q$. We note that $Q$ is a constant value, which leads to the outcome that the beamformer is $h_d^{\dag}/|h_d|$ and the transmit power varies as per $|h_u|^2$. In this CIPC, Eve cannot obtain any information on the eavesdropper's channel. In addition, the varying $P_t$ increases the uncertainty on the eavesdropper's channel, which can further enhance the physical layer security. Secrecy performance analysis on this CIPC and the optimization of $Q$ subject to a maximum transmit power constraint are challenging problems, since the communication from Alice to Eve is non-coherent with a random transmit power. {We note that there is a limitation on this CIPC caused by the required channel reciprocity: this CIPC can only be used in time division duplex (TDD) systems since channel reciprocity normally does not exist in frequency division duplex (FDD) systems. In addition, the performance of this CIPC will be affected by the non-perfect channel reciprocity in TDD systems and therefore channel reciprocity calibration should be adopted to counteract this lack of full reciprocity.}

\subsection{Location-Based Beamforming with and without Artificial Noise}

Multi-antenna techniques (e.g., beamforming) are desirable for enhancing reliability and security in wireless communications. Such techniques require accurate CSI to achieve the expected performance gain. However, in URLLC it is hard or infeasible to obtain accurate CSI due to the ultra-low latency requirement, since determining a complex channel matrix costs a large number of channel uses (especially when the antenna number is large). Against this background, we find that location-based beamforming can serve as an alternative solution to improve reliability and security in some URLLC scenarios where line-of-sight (LOS) components exist in the channel~\cite{yan2016location}. Such location-based beamforming, however, cannot
achieve the same level of performance gain as  traditional CSI-based beamforming, but can meet the ultra-low latency requirement in URLLC since it only requires the relative location information of the transceivers rather than a complex channel matrix. The accuracy of the location information and the weight of the LOS component determines whether (and how much) the use of artificial noise enhances the secrecy performance of the location-based beamforming.

\section{Conclusions}\label{sec:conclusion}

Due to its low complexity, physical layer security is able to provide security without violating ultra-low latency requirements, and  potentially serves as the main technique to enable security in URLLC. In this article, we identified the security gap and rate interval as two useful performance metrics for evaluating physical layer security in the context of URLLC. We also provided wider guidelines on analyzing and enhancing physical layer security in the context of URLLC.
Furthermore, we clarified critical issues related to the use of CSI in achieving physical layer security in URLLC, and presented potential techniques to overcome these issues.


\balance

\vspace{-0.5cm}

\begin{IEEEbiographynophoto}\\
Riqing Chen (riqing.chen@fafu.edu.cn) received the B.Eng. degree in communication engineering from Tongji University, China,
in 2001, the M.Sc. degree in communications and signal processing from Imperial College London,
U.K., in 2004, and the Ph.D. degree in engineering science from the University of Oxford, U.K.,
in 2010. Since 2014, he has been with the Institute of
the Cloud Computing and Big Data for Smart Agriculture and Forestry, Faculty of Computer Science
and Information Engineering, Fujian Agriculture and
Forestry University, China. His current research
interests include big data and visualization, cloud computing, consumer
electronics, and wireless communications.
\end{IEEEbiographynophoto}

\vspace{-1cm}

\begin{IEEEbiographynophoto}\\
Chunhui Li [S'17] (chunhui.li@anu.edu.au) received the B.S. degree in Automation and the M.S. degree in Digital Systems and Telecommunications from the Shanghai University of Engineering Science, China and the Australian National University in 2012 and 2015, respectively. He is currently a PhD student in the Research School of Electrical, Energy and Materials Engineering, the Australia National University, Canberra, Australia. His current research interests include ultra-reliable low-latency communications, physical-layer security, and unmanned aerial vehicle communications.
\end{IEEEbiographynophoto}

\vspace{-1cm}

\begin{IEEEbiographynophoto}\\
Shihao Yan [S'11-M'15] (shihao.yan@mq.edu.au) received the Ph.D. degree in Electrical Engineering from The University of New South Wales, Sydney, Australia, in 2015. From 2015 to 2017, he was a Postdoctoral Research Fellow in the Research School of Engineering, The Australian National University, Canberra, Australia. He is currently a University Research Fellow in the School of Engineering, Macquarie University, Sydney, Australia.
His current research interests are in the areas of wireless communications and statistical signal processing, including physical layer security, covert communications, and location spoofing detection.
\end{IEEEbiographynophoto}

\vspace{-1cm}

\begin{IEEEbiographynophoto}\\
Robert Malaney [M'02] (r.malaney@unsw.edu.au) is currently an Associate Professor in the School of Electrical Engineering and Telecommunications at the University of New South Wales (UNSW), Sydney, Australia. He holds a PhD in Physics from the University of St. Andrews, Scotland. Prior to joining UNSW, he was a NATO Fellow at the California Institute of Technology, a Research Fellow at the University of California Berkeley, a Senior Research Associate at the University of Toronto, and a Principal Research Scientist at CSIRO. Robert has over 150 peer-reviewed research publications in the areas of quantum mechanics, classical communications and quantum communications.
\end{IEEEbiographynophoto}

\vspace{-1cm}

\begin{IEEEbiographynophoto}\\
Jinhong Yuan [M'02--SM'11--F'16] (j.yuan@unsw.edu.au) received the B.E. and Ph.D. degrees in electronics engineering in 1991 and 1997, respectively. In 2000, he joined the School of Electrical Engineering and Telecommunications, University of New South Wales, Sydney, Australia, where he is a Professor and Head of Telecommunication Group with the School. He has published two books, five book chapters, over 300 papers in telecommunications journals and conference proceedings. He is an IEEE Fellow and currently serving as an Associate Editor for the \textsc{IEEE Transactions on Wireless Communications}. His current research interests include error control coding and information theory, communication theory, and wireless communications.
\end{IEEEbiographynophoto}

\end{document}